\documentclass{article}
\usepackage{ijcai20}

\usepackage{times}
\usepackage{soul}
\usepackage{url}
\usepackage[hidelinks]{hyperref}
\usepackage[utf8]{inputenc}
\usepackage[small]{caption}
\usepackage{graphicx}
\usepackage{amsmath}
\usepackage{amsthm}
\usepackage{booktabs}
\usepackage{algorithm}
\usepackage{algorithmic}
\usepackage{amsfonts}
\usepackage{multirow}
\usepackage{xcolor}
\title{SANST: A Self-Attentive Network for Next Point-of-Interest Recommendation}

\author{
Qianyu Guo\footnote{Contact Author}\and
Jianzhong Qi\\
\affiliations
The University of Melbourne, Australia\\
\emails
qianyu@student.unimelb.edu.au,
jianzhong.qi@unimelb.edu.au
}

\begin{document}

\maketitle

\begin{abstract}
   Next \emph{point-of-interest} (POI) recommendation  aims to offer suggestions on which POI to visit next, 
    given a user's POI visit history. This problem has  a wide application in the tourism industry, and it is gaining 
    an increasing interest as more POI check-in data become available.  
    The problem is often modeled as a sequential recommendation problem to take advantage of the sequential patterns  
    of user check-ins, e.g., people tend to visit Central Park after The Metropolitan Museum of Art in New York City. 
    Recently, \emph{self-attentive networks} have been shown to be both effective and efficient 
    in general sequential recommendation problems, e.g., to recommend products, video games, or movies. 
    Directly adopting  self-attentive networks for next POI recommendation, however, 
    may produce sub-optimal recommendations.  
    This is because vanilla self-attentive networks do not consider 
    the spatial and temporal patterns of  user check-ins, which are two critical 
    features in next POI recommendation. To address this limitation, in this paper, 
     we propose a model named \emph{SANST} that incorporates spatio-temporal patterns 
     of user check-ins into self-attentive networks. To incorporate the spatial patterns, 
     we encode the relative positions of POIs into their embeddings before feeding the embeddings into 
     the self-attentive network. To incorporate the temporal patterns, 
     we discretize the time of POI check-ins and model the temporal relationship between 
      POI check-ins by a relation-aware self-attention module. 
     We evaluate the performance of our SANST model with three real-world datasets. The  results show that 
     SANST consistently outperforms the state-of-the-art  models, and the advantage in nDCG@10 is up to 13.65\%.
\end{abstract}

\section{Introduction}

Travel and tourism is a trillion dollar industry worldwide~\cite{statista}.
To improve travel and tourism experiences, many location-based services are built. 
\emph{Next POI recommendation} is one of such services that is gaining 
    an increasing interest as more POI check-in data become available.  
Next POI recommendation aims to  suggest a POI to visit next, 
    given a user's POI visit history. Such a service is beneficial to both 
 the users and the tourism industry, since it  alleviates the burden of travel planning for the users while also boosts the visibility of POIs for the tourism industry.

 \begin{figure}[t]
    \centering
    \includegraphics[width=0.8\linewidth]{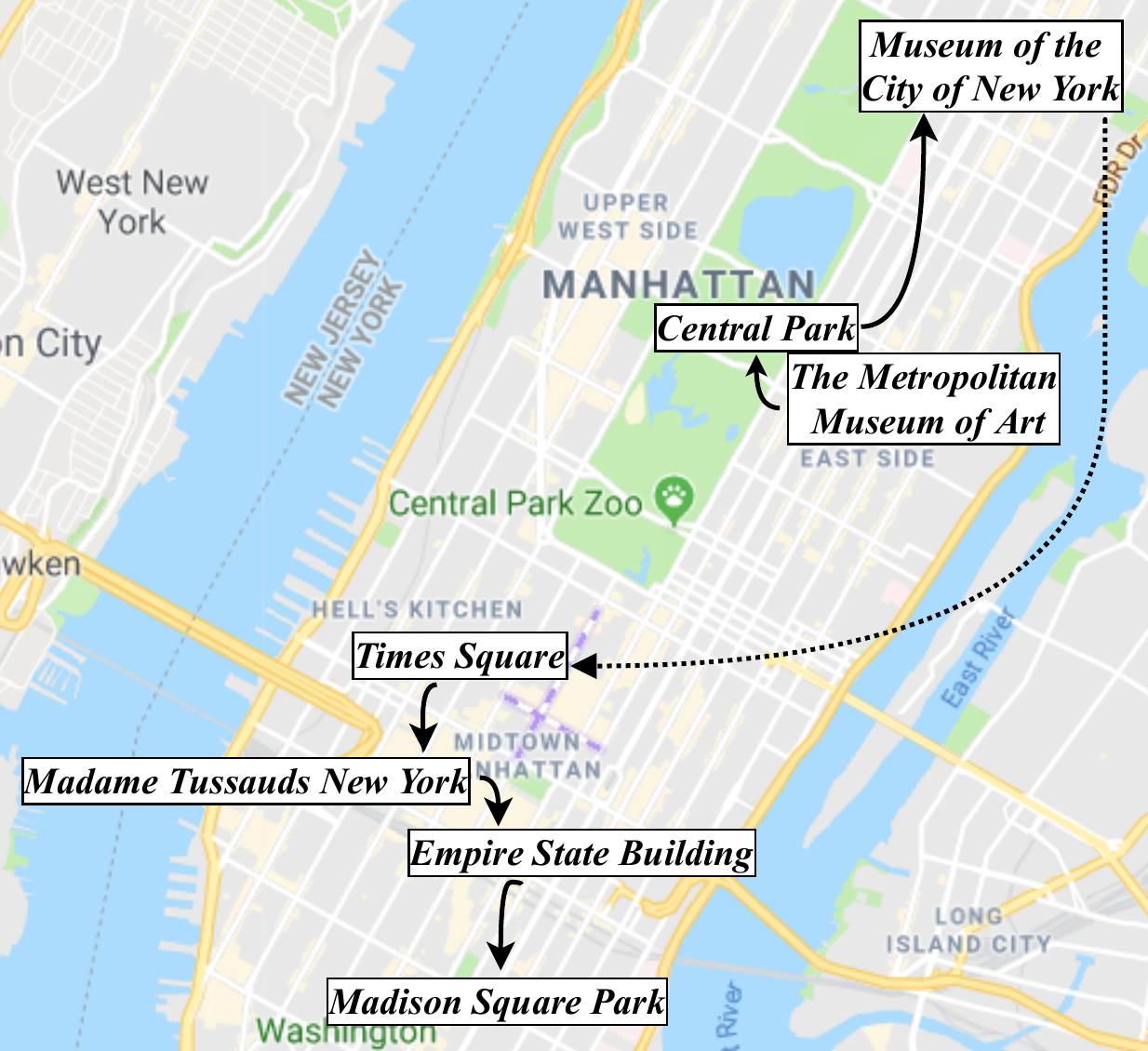}
    \caption{A POI check-in sequence in New York City (dashed line indicates movements across days)}
    \label{fig:sequence_example}
\end{figure}

 Next POI recommendation is often modeled as a \emph{sequential recommendation} problem~\cite{cheng2013you,feng2015personalized,feng2017poi2vec}, 
 to take advantage of the sequential patterns in  POI visits. 
 For example, tourists at New York City often visit Central Park right after The Metropolitan Museum of Art (``the Met'', cf.~Fig.~\ref{fig:sequence_example}). 
 If a user has just checked-in at the Met, Central Park is the next POI to recommend. 
 
 Recently, \emph{self-attentive networks} (SAN)~\cite{vaswani2017attention}, a highly effective and efficient  \emph{sequence-to-sequence} learning model, 
 have been introduced to \emph{general} sequential recommendation problems. 
The resultant model named  \emph{SASRec}~\cite{Kang2018Self} 
 yields state-of-the-art  performance in recommending next product or video game to purchase, next movie to watch, etc.
 
  Applying SASRec to make  next POI recommendations directly, however, may produce sub-optimal results. 
  This is because SASRec is designed for general recommendation scenarios and focuses  only on the sequential patterns in the input sequence. It does not consider any spatial or temporal patterns, 
  which are inherent in POI visit sequences and are critical for POI recommendations.  
  
  In terms of spatial patterns, as illustrates in Fig.~\ref{fig:sequence_example},  POI visits demonstrate a clustering effect~\cite{cheng2013you}. POIs nearby have a much larger probability to be visited consecutively 
 than those far away. This offers an important opportunity to alleviate the data sparsity problem resulted from 
   relying only on historical check-in sequences. In an extreme case, we may recommend  Central Park to a user at the Met, even if 
   the two POIs had not appeared in the historical check-in sequences, since both POIs are just next to each other and may  be visited  together. 
   
   In terms of temporal patterns,  historical check-ins made at different times ago shall have different impact on the next POI visit. 
  For example, in Fig.~\ref{fig:sequence_example}, the solid arrows represent transitions made within the same day, while the dashed arrow represents 
 a transition made  across days. Check-ins at Central Park and the Met may have a strong impact on the check-in at Museum of the New York City, as they together 
 form a day trip. In contrast, the check-in at Museum of the New York City may have little impact on the check-in at Times  Square, as they are at different days and may well
  be different trips. SASRec ignores such time differences and just focuses 
   on the transition probabilities between actions (i.e., check-ins). 
In SASRec, the impact of the check-in at the Met on the check-in at Central Park may be the same 
as that of the check-in at  Museum of the New York City on the check-in at Times  Square. 
Such an impact modeling strategy may be  inaccurate for next POI recommendations.    
   
To address these limitations, 
in this paper, we introduce self-attentive networks to next POI recommendation and integrate 
 \underline{s}patial and \underline{t}emporal pattern learning into the model.  
 We name our resultant model the \emph{SANST} model.  
 
 To integrate  spatial pattern learning, we learn a spatial embedding for each POI 
 and use it as the input of the self-attentive networks. Our embedding preserves 
 the spatial proximity between the POIs, such that 
 the self-attentive networks can learn not only the \emph{sequential} patterns between the check-ins but also the 
 \emph{spatial} patterns between the check-ins. 
 To learn our spatial embedding, we first 
 hash the POIs into a grid where  nearby cells are encoded by strings with common prefixes (e.g., following a \emph{space-filling curve} such as the \emph{Z-curve}).  
 We then learn character embeddings from the hashed strings corresponding to the POIs using 
 \emph{Bi-LSTM}~\cite{hochreiter1997long}.  The Bi-LSTM output is then used as the embedding of the POI.
   Since POIs at nearby cells are encoded by similar strings, they are expected to obtain similar embeddings in our model. Thus, we preserve the spatial proximity patterns 
   in POI check-ins. 
 
To integrate  temporal pattern learning, we 
follow~\cite{shaw2018self} to adapt the attention mechanism. 
We add a parameter to represent the relative position between two 
input elements $s^u_i$ and $s^u_j$ (i.e., check-ins) in the input sequence. 
We define the relative position based on the time difference between 
$s^u_i$ and $s^u_j$ instead of the number of other check-ins between $s^u_i$ and $s^u_j$ (which was done by~\cite{shaw2018self}). 
This way, our SANST model  represents the temporal patterns in check-ins explicitly
and can better learn their impact.
    
To summarize, we make the following contributions:
\begin{enumerate}
\item We propose a self-attentive network named SANST that incorporates spatial and temporal POI check-in patterns for next POI recommendation. 
    \item To incorporate the spatial patterns of POI check-ins, we propose a novel POI embedding technique that 
    preserves the spatial proximity of the POIs. Our technique hashes POIs into a grid where  nearby cells are encoded by strings with common prefixes.
    The POI embeddings are learned from the hashed strings via character embeddings, such that POIs at nearby cells yield similar embeddings.
   
    \item To incorporate the temporal patterns of POI check-ins, we extend the self-attentive network 
     by adding a parameter to represent the relative time between the check-ins. This enables the network 
     to learn the temporal patterns of the check-ins explicitly. 
    
    \item We study the empirical performance of our SANST model  on three real-world datasets. 
    The  results show that SANST outperforms state-of-the-art sequential next POI recommendation models and adapted models that combine self-attentive networks with spatial feature learning directly by up to 13.65\% in terms of nDCG@10.
\end{enumerate}

\section{Related Work}

Like many other recommendation problems, POI recommendation has attracted extensive research interests. 
Earlier  studies on this topic use \emph{collaborative filtering} (CF) techniques, including both \emph{user-based CF} (UCF)~\cite{ye2011exploiting} 
and \emph{item-based CF} (ICF)~\cite{levandoski2012lars}. These techniques make recommendations based on either user or item similarity.
\emph{Factorization models}~\cite{gao2013exploring,li2015rank,Lian:2014:GJG:2623330.2623638} are also widely used, 
where the user-POI matrix is factorized to learn users' latent interests towards the POIs.
Another stream of studies use probabilistic models~\cite{cheng2012fused,kurashima2013geo}, which 
aim to model the mutual impact between spatial features and user interests (e.g., via Gaussian or topic models).
Details of these studies can be found in a survey~\cite{yu2015survey} and 
 an experimental paper~\cite{Liu:2017:EEP:3115404.3115407}.

\textbf{Next POI recommendation.}
In this paper, we are interested in a variant of the POI recommendation problem, i.e., \emph{next POI recommendation} (a.k.a. successive POI recommendation). 
This variant aims to recommend the very next POI for a user to visit, given the user's past POI check-in history as a sequence. 
Users' sequential check-in patterns play a significant role in this problem. 
For example, a tensor-based model named \emph{FPMC-LR}~\cite{cheng2013you} recommends the next POI by considering the successive POI check-in relationships.
It extends the \emph{factorized personalized Markov chain} (FPMC) model~\cite{Rendle:2010:FPM:1772690.1772773}  by factorizing the transition probability with users' movement constraints.
Another model named \emph{PRME}~\cite{feng2015personalized} takes a \emph{metric embedding} based approach to learn sequential patterns and individual preferences. \emph{PRME-G}~\cite{feng2015personalized}  further incorporates spatial influence using a weight function based on spatial distance.
\emph{POI2Vec}~\cite{feng2017poi2vec}  also makes recommendations based on user and POI embeddings. To learn the embeddings, it adopts 
the \emph{word2vec} model~\cite{DBLP:journals/corr/abs-1301-3781} originated from the natural language processing (NLP) community; users' past POI check-in sequences form the ``word contexts'' 
for training the word2vec model. 
The POI check-in time is also an important factor that is considered in next POI recommendation models. 
For example, \emph{STELLAR}~\cite{zhao2016stellar} uses \emph{ranking-based pairwise tensor factorization} to model the interactions among users, POIs, and time. 
\emph{ST-RNN}~\cite{liu2016predicting} extends \emph{recurrent neural networks} (RNN) to incorporate both spatial and temporal features by adding distance-specific and time-specific transition matrices into 
the model state computation. 
\emph{MTCA}~\cite{li2018next} and \emph{STGN}~\cite{zhao2019go} adopt LSTM based models to capture the spatio-temporal information.  \cite{yuan2013time} split a day into time slots (e.g., by hour) to learn the periodic temporal patterns of POI check-ins.
 \emph{LSTPM}~\cite{sun2020lstpm} map a week into time slots. They propose a context-aware long and short-term preference modeling framework to model users' preferences and a geo-dilated RNN to model the non-consecutive geographical relation between POIs.

\textbf{Attention networks for recommender systems.}
Due to its high effectiveness and efficiency, the \emph{self-attention} mechanism~\cite{vaswani2017attention} has been applied to various tasks such as machine translation. 
The task of recommendation makes no exception. \emph{SASRec}~\cite{Kang2018Self} is a  sequential recommendation model based on self-attention. It extracts 
the historical item sequences of each user, and then maps the recommendation problem to a sequence-to-sequence learning problem. 
\emph{AttRec}~\cite{zhang2018next} uses self-attention to learn from  user-item interaction records about their recent interests, which 
are combined with users' long term interests learned by a metric learning component to make recommendations. 
These models have shown promising results in general sequential recommendation problems, e.g., to recommend products, video games, or movies.
However, they are not designed for POI recommendations and have not considered the spatio-temporal patterns in POI recommendations. 
In this paper, we build a self-attentive network based on the SASRec model to  incorporate spatio-temporal patterns of user check-ins. We will compare 
with SASRec in our empirical study. We omit AttRec since SASRec has shown a better result. 

\begin{table}[t]
\centering
\begin{small}
\setlength{\belowcaptionskip}{5pt}%
\caption{Frequently Used Symbols}
\begin{tabular}{llcc}  
\toprule
Symbol & Description\\ \hline 
\midrule
$U$ & A set of users \\         
$L$ & A set of POIs \\
$T$ & A set of check-in time \\
$S^{u}$ & The historical check-in sequence of a user $u$ \\
$s^{u}_i$ & A historical check-in of a user $u$ \\
$\ell$& The maximum  check-in sequence length\\
$\mathbf{E}$& The input embedding matrix\\
$r_{l, i}$& A relevant score computed by SASRec\\
 \bottomrule
\end{tabular}
\label{tab:effect_num_headers}
 \end{small}
\end{table}

\begin{figure*}[t]
    \begin{center}
        \includegraphics[width=.99\textwidth]{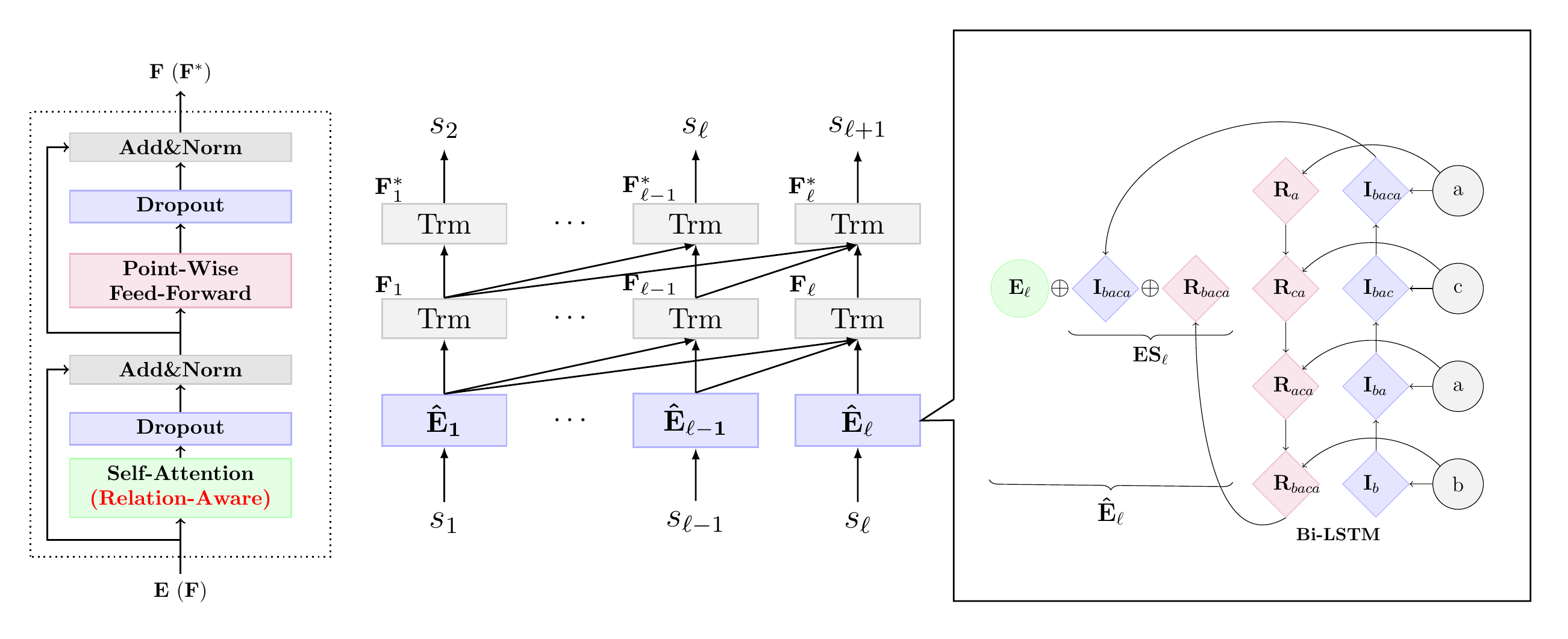}
    \end{center}
    \footnotesize \hspace{.25cm} (a) Transformer layer (Trm) \hspace{2cm} (b) SASRec model structure \hspace{3.25cm} (c) Grid cell ID string embedding
    \caption{The architecture of our SANST model}
    \label{fig:architecture}
\end{figure*}

\section{Preliminaries}
We start with basic concepts and a problem definition in this section. 
We then present the \emph{SASRec} model~\cite{Kang2018Self}, based on which our proposed model is built. 
This model will also be used in our experiments as a baseline. We summarize the frequently used symbols 
 in Table~\ref{tab:effect_num_headers}.

\subsection{Problem Definition}
We consider a set of users $U$ and a set of POIs $L$. 
Each user $u \in U$ 
comes with a historical POI check-in sequence $S^u$ = $\langle s_1^u, s_2^u, ..., s_{|S^u|}^u \rangle$, 
which is sorted in ascending order of time.  Here, $|S^u|$ denotes the size of the set $S^u$. 
Every check-in $s_i^u \in S^u$ is a tuple $\langle s_i^u.l, s_i^u.t\rangle$, where $s_i^u.l \in L$ 
is the check-in POI and $s_i^u.t$ is the check-in time, respectively. 

Given a user $u\in U$ and a time $t_q$, our aim is to predict $s_{|S_u|+1}^u.l \in L$, i.e., the next POI that $u$ will visit at $t_q$. 
In our model, we consider check-in times in the day granularity, to alleviate the data sparsity issue.



\subsection{SASRec}\label{sec:sasrec}

SASRec is a  two-layer  \emph{transformer} network~\cite{vaswani2017attention}. 
It models sequential recommendation as a sequence-to-sequence learning problem that translates 
an input sequence $\langle s_1, s_2, ..., s_\ell\rangle$ to an output sequence $\langle s_2, s_3, ..., s_{\ell+1}\rangle$. 
Here, $\ell$ is a hyperparameter  controlling the input length. 
The last element in the output, $s_{\ell+1}$, is the recommendation output.  
SASRec can be adopted for next POI recommendation directly, by treating a user's check-in sequence $S^u = \langle s^u_1, s^u_2, ..., s^u_{|S^u|} \rangle$
as the model input. If $|S^u| > \ell$, only the latest 
$\ell$ check-ins are kept; if $|S^u| < \ell$, the input sequence is padded by 0's at front. 

Fig.~\ref{fig:architecture}b illustrates the SASRec model structure. 
The input sequence of SASRec first goes through an embedding layer to convert 
every input element $s_i$ (e.g., a POI id $s^u_i.l$) into a $d$-dimensional vector $\mathbf{E_i} \in \mathbb{R}^d$.

The embeddings are then fed into two stacking transformer layers. 
In the first transformer layer, as shown in Fig.~\ref{fig:architecture}a, the transformer input 
first goes through a self-attention layer to capture the
pairwise dependency between the POI check-ins. 
Specifically, the embeddings of the input elements are  concatenated rowwise to form a matrix $\mathbf{E} = [\mathbf{E_1}^{T}; \mathbf{E_2}^{T}; ... \mathbf{E_\ell}^{T}]^{T}$. 
Then, three linear projections on  $\mathbf{E}$ are done using three projection matrices 
$\mathbf{W^{Q}} , \mathbf{W^{K}} , \mathbf{W^{V}} \in \mathbb{R}^{d\times d}$, respectively. 
These matrices will be learned by the model. The linear projections yield three matrices $\mathbf{Q} = \mathbf{E}\mathbf{W^Q}$, 
$\mathbf{K} = \mathbf{E}\mathbf{W^K}$, and $\mathbf{V} = \mathbf{E}\mathbf{W^V}$, respectively. 
The self-attention for $\mathbf{E}$, denoted by $sa(\mathbf{E})$, is then computed as follows, 
where $\mathbf{Q}$, $\mathbf{K}$, and $\mathbf{V}$ are let be the same: 
\begin{equation}
sa(\mathbf{E})  = \textrm{softmax}(\frac{\mathbf{QK}^T}{\sqrt{d}}) \mathbf{V}
\end{equation}
To endow the model with non-linearity,  a \emph{point-wise feed-forward network} is applied 
on $sa(\mathbf{E})$. Let $\mathbf{S_i}$  be the $i$-th output of of self-attention module. Then, 
the  feed-forward network on $\mathbf{S_i}$ is computed as: 
\begin{equation}
\mathbf{F_{i}}= \textrm{ReLU}(\mathbf{S_{i}W}^{(1)}+\mathbf{b}^{(1)})\mathbf{W}^{(2)}+\mathbf{b}^{(2)}
\end{equation}
Here, $\mathbf{W}^{(1)}, \mathbf{W}^{(2)} \in \mathbb{R}^{d\times d}$ and 
$\mathbf{b}^{(1)}, \mathbf{b}^{(2)} \in \mathbb{R}^{d\times 1}$ are learnable parameters. 
Layer normalization and dropout are adopted in between these layers to avoid overfitting.

Two transformer layers are stacked to form a deeper network, where the 
point-wise feed-forward network of the first  layer is used as the input 
as the second layer. 

Finally, the prediction output at position $i$ is produced by computing the relevance score $r_{l, i}$ between 
a POI $l$ and the position-$i$ output $\mathbf{F^*_{i}}$ of the point-wise feed-forward network of the second transformer layer. 
\begin{equation}
r_{l,i} = \mathbf{F^*_{i}} \mathbf{E}(l)^T
\end{equation}
Here, $\mathbf{E}(l)$ denotes the embedding of $l$ fetched from the embedding layer. 
The POIs with the highest scores is returned. 

To train the model, the binary cross-entropy loss is used as the objective function:
\begin{equation}\label{eq:loss}
-\sum_{u \in U} \sum_{i=1}^{\ell} \Bigg[ \textrm{log}(\sigma(r_{s^u_i.l,i})) +\sum_{l \notin S^{u}} \textrm{log}(1-\sigma(r_{l,i}))\Bigg]
\end{equation}

\section{Proposed Model}

As discussed earlier, while SASRec can be adapted to 
make  next POI recommendations, a direct  adaptation may produce sub-optimal results. 
  This is because SASRec does not consider any spatial or temporal patterns, 
  which are inherent in POI visit sequences and are critical for POI recommendations.  
   In this section, we present our \emph{SANST} model to address this  limitation 
   via   incorporating spatial and temporal pattern learning into self-attentive networks.
   As illustrated in Fig.~\ref{fig:architecture}, our SANST model shares the overall structure with SASRec. We detail below how to  incorporate spatial and temporal pattern learning into this structure. 

\subsection{Spatial Pattern Learning}
To enable SANST to learn the spatial patterns in POI check-in transitions, 
we update the embedding $\mathbf{E_i}$ of the $i$-th  check-in 
of an input sequence to incorporate the location of the checked-in POI.  
This way, our SANST model can learn not only the transitions between the POIs (i.e., their IDs) but 
also the transitions between their locations.

A straightforward approach to incorporate the POI locations is to concatenate the geo-coordinates 
of the POIs  with the SASRec embedding $\mathbf{E_i}$.
This approach, however, suffers from the data sparsity problem -- geo-coordinates of POIs are in an infinite and continuous 
space, while the number of POIs is usually limited (e.g., thousands).  

To overcome the data sparsity, we discritize the data space with a grid such that POI locations are represented by 
the grid cell IDs. We learn embeddings for the grid cell IDs (detailed next). Then, given a check-in $s^u_i$, 
we locate the grid cell in which the check-in POI $s^u_i.l$ lies. 
We use the grid cell ID embedding as the spatial embedding of $s^u_i$, denoted by $\mathbf{ES_i}$. 
We concatenate (denoted by $\oplus$)  $\mathbf{ES_i}$ with 
the SASRec embedding $\mathbf{E_i}$ to form a new  POI embedding in our SANST model, denoted by $\hat{\mathbf{E_i}}$: 
\begin{equation}
\hat{\mathbf{E_i}} = \mathbf{E_i} \oplus \mathbf{ES_i}
\end{equation}

\subsubsection{Grid Partitioning and Grid Cell Encoding}
We use a \emph{space-filling curve} to partition the data space 
and to number the grid cells. We encode the grid cell numbers 
 into strings and use the encoded strings as the grid cell IDs.
The purpose is to generate ID strings such that nearby grid cells 
have similar ID strings. This way, the cell IDs can preserve the spatial 
proximity of the corresponding cells. We adapt the \emph{GeoHash} technique to generate the strings as detailed below.\footnote{https://en.wikipedia.org/wiki/Geohash}

\begin{figure}[t]
\begin{center}
\includegraphics[width=0.95\columnwidth]{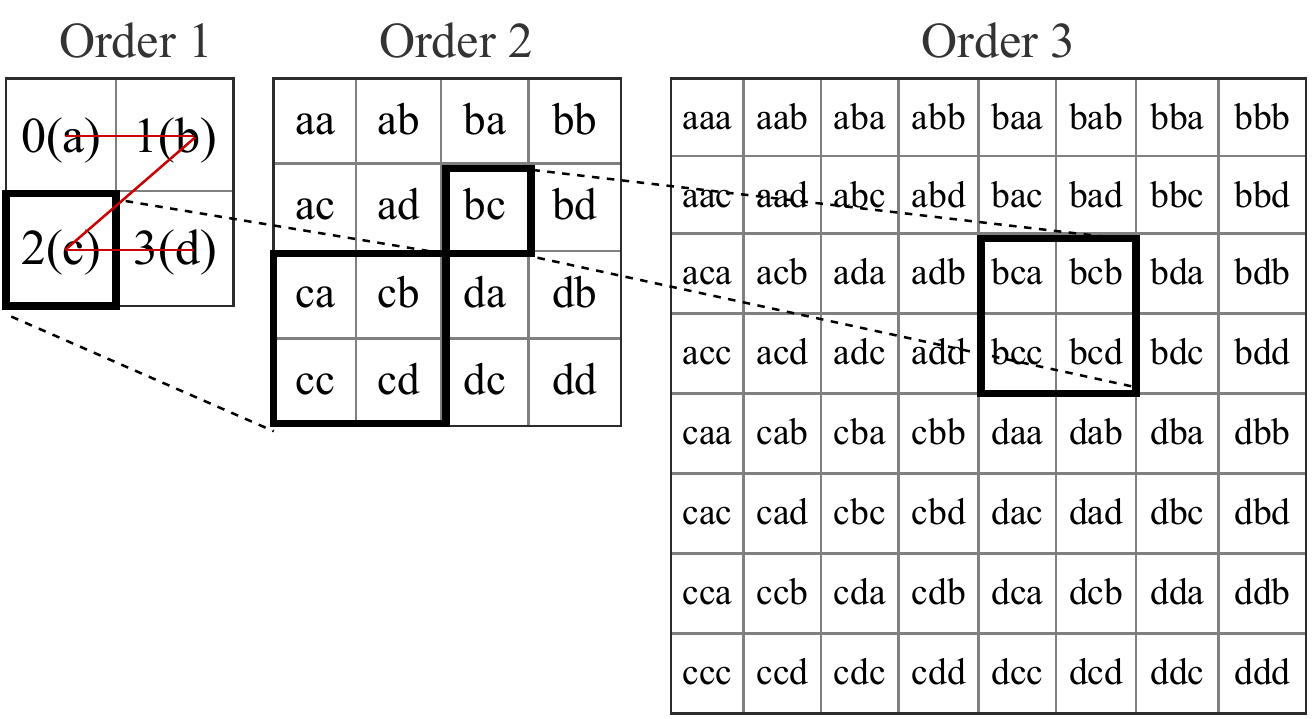}
\caption{Example of grid cell encoding}
\label{fig:geohash_overview}
\end{center}
\end{figure}

Fig.~\ref{fig:geohash_overview} illustrates the steps of 
the grid partitioning and grid cell encoding scheme. A \emph{Z-curve} is used 
in this figure, although other space-filling curves such as \emph{Hilbert-curves} may be used as well. 
Suppose that an order-$n$ curve is used to partition the grid.  Then, there are 
$2^n \times 2^n$ grid cells, and each curve value is represented by an $2n$-bit integer. 
We break a curve value into segments of $\gamma$ bits from the left to the right consecutively (a total of $\lceil n/\gamma \rceil$ segments). Each segment is then 
mapped to a character in a size-$2^\gamma$ alphabet. In the figure, we use $\gamma = 2$ and an alphabet of $2^\gamma = 4$ characters (i.e., `a', `b', `c', and `d'). 
As the figure shows, using this encoding, nearby grid cells obtain similar ID strings -- 
they share common prefixes by the definition of the curve values. The longer the common prefix is, the nearer the two cells are. 
In our experiments, we use $n=60, \gamma=5$, and $\lceil n/\gamma \rceil = 12$ .

\subsubsection{Grid Cell ID Embedding Learning}

Geohash encodes 
coordinates in a hierarchical way. It can create adjacent cells without a common prefix, e.g., cells ‘cdb’ and ‘dca’ in Fig.~\ref{fig:geohash_overview}. 
We address this problem by a natural language processing approach -- we learn 
an ID string embedding via learning character embeddings using a Bi-LSTM network.  
As Fig.~\ref{fig:architecture}c shows, each character (a  randomly initialized $d_s$-dimensional vector to be learned) of an ID string (e.g., ``baca'') is fed into Bi-LSTM. The final output of  
Bi-LSTM  in both directions are caught (i.e., $\mathbf{I}_{baca}$ and $\mathbf{R}_{baca}$), which are used as the ID string embedding (i.e., spatial embedding $\mathbf{ES_i}$) to form our POI check-in embedding 
$\hat{\mathbf{E_i}}$ in SANST. 
Since the character embeddings are trained jointly with the entire model, the characters which are adjacent to each other in the grid cells will have similar weights in there embedding vectors. Such as ‘a’ and ‘b’, ‘a’ and ‘c’. Therefore, for the cells labeled with  ‘cbd’ and ‘dca’, even though they do not share a common prefix, the adjacency relation between the characters in each layer (e.g., ‘c’ and ‘d’ in the top layer of the hierarchical hash codes) can still be captured. We envision that this spatial embedding method can not only be used in our problem but also other tasks that incorporate spatial information.


\subsection{Temporal Pattern Learning}

To learn the temporal patterns in POI check-in transitions, 
we follow~\cite{shaw2018self} to adapt the attention mechanism. 
We add a parameter to represent the relative position between two 
 check-ins $s^u_i$ and $s^u_j$ in the input sequence. 
We define the relative position based on the time difference between 
$s^u_i$ and $s^u_j$ instead of the number of other check-ins between $s^u_i$ and $s^u_j$ (which was done by~\cite{shaw2018self}). 
This way, we model  the temporal patterns explicitly
and can better learn their impact. We detail our adaptation next. 

In self-attention, each output element $\mathbf{S_i}$ is computed as a weighted sum of linearly transformed input elements, i.e., 
$\mathbf{S_{i}} = \sum_{j=1}^{\ell}\alpha_{ij} (\mathbf{x_{j}}\mathbf{W^{V}})$, where $\mathbf{x_{j}}$ denotes the position-$i$ input element (e.g., a POI check-in).
\cite{shaw2018self} add an edge between two input elements $\mathbf{x_{i}}$ and $\mathbf{x_{j}}$ to model their relative position
in the input sequence. The impact of this edge is learned from two  
vectors $\mathbf{a_{ij}^{V}}$ and $\mathbf{a_{ij}^{K}}$, and the self-attention equation is updated to: 
\begin{equation}
\mathbf{S_{i}} = \sum_{j=1}^{\ell}\alpha_{ij} (\mathbf{x_{j}}\mathbf{W^{V}}+\mathbf{a_{ij}^{V}})
\end{equation}
Here,  $\alpha_{ij}$ is computed as: 
\begin{equation}
\alpha_{ij} = \text{softmax}(\frac{\mathbf{x_{i}W^{Q}}(\mathbf{x_{j}W^{K}}+\mathbf{a_{ij}^{K}})^{T}}{\sqrt{d}})
\end{equation}
We adapt $\mathbf{a_{ij}^{K}}$ and $\mathbf{a_{ij}^{V}}$ as follows to learn the temporal pattern between 
$\mathbf{x_{i}}$ and $\mathbf{x_{j}}$ (i.e., the relative position in time): 
\begin{gather}
    \mathbf{a_{ij}^{K}} = \mathbf{w^{K}_{clip(T_{j}-T_{i},k)}} \\
    \mathbf{a_{ij}^{V}} = \mathbf{w^{V}_{clip(T_{j}-T_{i},k)}} \\
    clip(x,k) = \max\{-k,\min\{k,x\}\}
\end{gather}

\begin{figure}[t]
\begin{center}
\includegraphics[width=0.95\columnwidth]{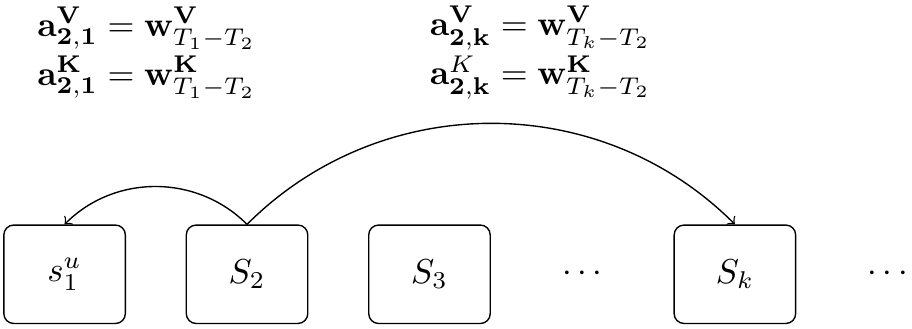}
\caption{Temporal pattern learning}
\label{fig:time_overview}
\end{center}
\end{figure}

Here, $T_{i}$ and $T_{j}$ represent the temporal label on input sequence at position ${i}$ and ${j}$, respectively. 
We compute the temporal label for the $i$-th input element of user $u$ as $T_i =(t_q - s_i^u.t)$.  We then learn relative temporal representations $\mathbf{w^{K}}=(\mathbf{w^{K}_{-k}},...,\mathbf{w^{K}_{k}})$ and $\mathbf{w^{V}}=(\mathbf{w^{V}_{-k}},...,
\mathbf{w^{V}_{k}})$, where $k$ is a hyperparameter that represents the size of the time context window that we examine.
We illustrate these concepts in Fig.~\ref{fig:time_overview}, where each input element is a user check-in $s_i^u$.

\subsection{Model Training}
We use the same loss function as that of SASRec (i.e., Equation~\ref{eq:loss}) to train SANST. We randomly generate one
negative POI for each time step in each sequence in each epoch. 
The model is optimized by the Adam optimizer.

\section{Experiments}
We perform an empirical study on the proposed model SANST and compare it with state-of-the-art
sequential recommendation and next POI recommendation models.

\begin{table}[t]
\centering  
\small
\setlength{\belowcaptionskip}{5pt}%
\setlength{\tabcolsep}{2pt}
\caption{Dataset Statistics}\
\begin{tabular}{lllll}  
\toprule
Dataset &$\#$ users  & $\#$ POIs &  $\#$ check-ins & Time range\\
\midrule
Gowalla     &	10,162	&   24,250	&   456,988 & 02/2009-10/2010\\
Los Angeles &    3,202  &    4,368  &   101,327 & 03/2009-10/2010\\         
Singapore   &	 2,321	&    5,596  &   194,108 & 08/2010-07/2011\\    
\bottomrule
\end{tabular}
\label{tab:dataset}
\end{table}
\renewcommand{\arraystretch}{1} 

\begin{table*}[t]
\centering
\small
\selectfont
  \caption{Summary of Results}
  \label{tab:performance_comparison}
    \begin{tabular}{clcccccc}
    \toprule & 
    \multirow{2}{*}{Method} &
    \multicolumn{2}{c}{Gowalla} & \multicolumn{2}{c}{Los Angeles} &  \multicolumn{2}{c}{Singapore}
    \cr \cmidrule(lr){3-4} \cmidrule(lr){5-6}\cmidrule(lr){7-8} &&hit@10&nDCG@10&hit@10&nDCG@10&hit@10&nDCG@10\cr
    \midrule
    \multirow{6}{*}{Baseline} &
    FPMC-LR      &   0.1197  &   0.0741  &    0.2347  &   0.1587  &   0.1784  &   0.1017   \cr &   
    POI2vec      &   0.0939  &   0.0606  &    0.2370  &   0.1700  &   0.2063  &   0.1425   \cr &
    PRME-G       &   0.1852  &   0.1083  &    0.2367  &   0.1592  &   0.1601  &   0.1049   \cr &
    
    SASRec       &   0.2023  &   0.1209  &   0.3648   &   0.2337  &   0.2245  &   0.1429  \cr & 
    SASRec+WF    &   0.2096  &   0.1184  &   0.3304   &   0.2006  &   0.2137  &   0.1251 \cr &
    SASRec+2KDE  &   0.1842  &   0.1113  &   0.3057   &   0.1912  &   0.1900  &   0.1154  \cr
    &
    LSTPM  &   0.1361  &   0.0847  &   0.2366   &  0.1580  &   0.1777  &  0.1105  \cr
    \midrule
    \multirow{2}{*}{Variants}& 
    SANS         &   0.2248  &   0.1372  &    0.3891  &   0.2519  &   0.2417  &   0.1491  \cr & 
    SANT         &   0.2028  &   0.1245  &    0.3635  &   0.2264  &   0.2296  &   0.1441  \cr
    \midrule
    \textbf{Proposed} & 
    \textbf{SANST}&{\bf0.2273}&{\bf0.1374}&{\bf0.3941}&{\bf0.2558}&{\bf0.2417}&{\bf0.1531}\cr
        &      &(+8.44\%)&(+13.65\%)&(+8.03\%)&(+9.46\%)&(+7.66\%)&(+7.14\%)  \cr
    \bottomrule
    \end{tabular}
\end{table*}

\subsection{Settings}

We first describe our experimental setup, including datasets, baseline models, and 
model implementation details. 

\textbf{Datasets.} We evaluate the models on three real-world  datasets:   the \textbf{Gowalla} dataset~\cite{yuan2013time},  the \textbf{Los Angeles} dataset~\cite{cho2011friendship},
and the \textbf{Singapore} dataset~\cite{yuan2013time}. Table~\ref{tab:dataset} summarizes the dataset statistics. 
Following previous studies~\cite{feng2015personalized,cui2019distance2pre}, for each user's check-in sequence, we take the last POI for  testing and the rest for training,  and we omit users with fewer than five check-ins.

\textbf{Baseline models.}
We compare with six baseline models:
\begin{itemize}
\item \textbf{FPMC-LR}~\cite{cheng2013you}: This is a matrix factorization model that extends the factorized personalized Markov chain (FPMC) model~\cite{Rendle:2010:FPM:1772690.1772773}  by factorizing the POI transition probability with users' movement constraints.

\item \textbf{PRME-G}~\cite{feng2015personalized}: This is a metric embedding based approach 
to learn sequential patterns and individual preferences, and it incorporates spatial influence using a weight function based on spatial distance.

\item \textbf{POI2Vec}~\cite{feng2017poi2vec}:  This is an embedding based model. It adopts 
the word2vec model to compute POI and user embeddings. Recommendations are made based on the embedding similarity. 

\item \textbf{LSTPM}~\cite{sun2020lstpm}: This is an LSTM based model. It uses two LSTMs to capture users' long-term and short-term preferences and a geo-dilated RNN to model the non-consecutive geographical relation between POIs.

\item \textbf{SASRec}~\cite{Kang2018Self}:  This is the  state-of-the-art sequential recommendation model as described in the Preliminaries section.

\item \textbf{SASRec+WF}: We combine SASRec with spatial pattern learning by adopting a weight function~\cite{feng2015personalized} to weight the relevance score by the spatial distance to the last POI check-in. 

\item \textbf{SASRec+2KDE}: We combine SASRec with spatial pattern learning by adopting 
the \emph{two-dimensional kernel density estimation} (2KDE) model~\cite{zhang2014lore}. 
The 2KDE model learns a POI check-in distribution for a user based on the POI geo-coordinations.  
We weight the relevance score for a POI in SASRec  by the probability of the POI learned by 2KDE.
\end{itemize}

\textbf{Model variants.}
To study the contribution of the spatial and time pattern learning to the overall performance of SANST, 
we further compare with two model variants: \textbf{SANS} and \textbf{SANT}, which are 
SANST without time pattern learning and SANST without spatial pattern learning, respectively.

\textbf{Implementation details.}
For FPMC-LR, PRME-G, and POI2Vec, we use the code provided by~\cite{cui2019distance2pre} (we could not compare with the model proposed by~\cite{cui2019distance2pre} because they only provided implementations of their baseline models but not their proposed model). For LSTPM, we use the code provided by~\cite{sun2020lstpm}. For SASRec, we use the code provided by~\cite{Kang2018Self},
based on which our SANST model and its variants are implemented. 
The learning rate, regularization, POI embedding dimensionality, and batch size are set to 0.005, 0.001, 50, and 128 for all models, respectively. 
Other parameters of the baseline models are set to their default values that come with the original paper. 
For our SANST model and its variants, we use 2 transformer layers, a dropout rate of 0.3, a character embedding size $d_s$ of 20, and  the Adam optimizer. 
We set the maximum POI check-in sequence length $\ell$ to  100, and the time context window size $k$ to be 3, 5 and 1 for Gowalla, Los Angeles and Singapore,  respectively. We train the models with an Intel(R) Xeon(R) CPU @ 2.20GHz, a Tesla K80 GPU, and 12GB memory.

We report two metrics: \emph{hit@10} and \emph{nDCG@10}. They measure how likely the ground truth POI is in the top-10 
POIs recommended and the rank of the ground truth POI  in the top-10 POIs recommended, respectively.
\subsection{Results}

We first report comparison results with baselines and then 
report the impact of model components and parameters.

\textbf{Overall performance.}
Table~\ref{tab:performance_comparison} summarizes the model performance. 
We see that our model SANST outperforms all baseline models over all three datasets consistently, in terms of both hit@10 and nDCG@10. 
The numbers in parentheses show the improvements gained by SANST comparing with the best baselines. 
We see that SANST achieves up to 8.44\% and 13.65\% improvements in hit@10 and nDCG@10 (on Gowalla dataset), respectively. 
 These improvements are significant as confirmed by $t$-test with  $p<0.05$. Among the baselines, SASRec outperforms FPMC-LR,  POI2Vec
 PRME-G and LSTPM, which validates the effectiveness of self-attentive networks in making next item recommendations. However, adding spatial features to self-attentive networks with existing methods do not necessarily yield a better model. For example, both SASRec+WF and SASRec+2KDE produce worse results than SASRec for most cases tested. 
 They learn spatial patterns and sequential patterns separately which may not model the correlations between the two factors accurately. 
 Our spatial pattern learning technique differs from the existing ones in its ability to learn representations that preserve the spatial information 
and can be integrated with sequential pattern learning to form an integrated model. 
Thus, combining self-attentive networks with our spatial pattern learning technique (i.e., SANS) outperforms SASRec, while adding temporal pattern learning (i.e., SANST) further boosts our model performance. 

\textbf{Ablation study.} By examining the results of the two model variants SANS and SANT in Table~\ref{tab:performance_comparison}, 
we find that, while both spatial and temporal patterns may help next POI recommendation, spatial patterns 
appear to contribute more. We conjecture that this is due to the use of  the same  discretization (i.e., by day) 
across the check-in time of all users in SANT.  Such a global discretization method may not reflect the impact 
of time for each individual user accurately. It impinges the performance gain achieved from the time patterns. 
Another observation is that, when both types of patterns are used (i.e., SANST), the model achieves a higher accuracy comparing with 
using either pattern. This indicates that the two types of patterns  complement each other well. 

\begin{table}[t]
\setlength{\tabcolsep}{2pt}
\centering
\small
\caption{Impact of $\ell$ (NDCG@10)}
\label{tab:effect_sequence_length}
\begin{tabular}{lcccccc}
\toprule
Model  & \multicolumn{2}{c}{Gowalla} & \multicolumn{2}{c}{Los Angeles} & \multicolumn{2}{c}{Singapore} \cr
\cmidrule(lr){2-3} \cmidrule(lr){4-5} \cmidrule(lr){6-7}  
$\ell$  & SASRec & SANST & SASRec & SANST & SASRec & SANST \cr
\midrule
20   & 0.1131 & 0.1177 & 0.2240 & 0.2369 & 0.1289 & 0.1299 \cr
50   & 0.1265 & 0.1301 & 0.2326 & 0.2409 & 0.1379 & 0.1431 \cr
100  & 0.1209 & 0.1374 & 0.2337 & 0.2558 & 0.1429 & 0.1531 \cr
200  & 0.1251 & 0.1391 & 0.2195 & 0.2355 & 0.1446 & 0.1506 \cr
\bottomrule
\end{tabular}
\end{table}

Next, we study the impact of hyperparameters and model structure on the performance of  SANST. 
Due to space limit, we only report  results on nDCG@10  in these experiments. Results on hit@10 
show similar patterns and are omitted. 

\textbf{Impact of input sequence length $\ell$.}
We start with the impact of the input sequence length $\ell$. 
As shown in Table~\ref{tab:effect_sequence_length}, SANST outperforms the best baseline SASRec across 
all $\ell$ values tested. Focusing on our model SANST, as $\ell$ grows from 20 to 100, 
its nDCG@10 increases. This can be explained by that more information is available for SANST to learn the check-in patterns. 
When $\ell$ grows further, e.g., to 200, the model performance drops. This is because, on average, the user check-in sequences are much shorter 
than 200. Forcing a large input length requires padding 0's at front which do not contribute extra information. Meanwhile, 
looking too far back may introduce noisy check-in patterns.  Thus, a longer input sequence may not benefit the model.

\begin{figure}[t]
\centering
\includegraphics[width=.45\textwidth]{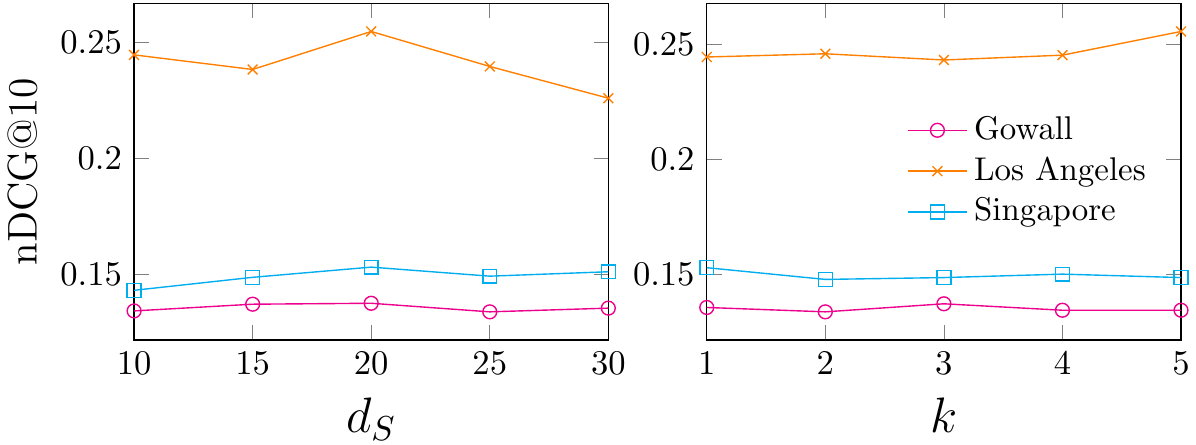} \\
(a) Character embedding size\ \ (b) Time window size
\caption{Impact of character embedding size $d_{S}$ and time context window size $k$}
\label{fig:ds_w}
\end{figure}

\textbf{Impact of character embedding dimensionality $d_s$.}
We vary the character embedding dimensionality $d_s$ for our spatial pattern learning module from 10 to 30 
and report the performance of SANST in Fig.~\ref{fig:ds_w}a. As the figure shows, our model 
is robust against the character embedding dimensionality (note the small value range in the $y$-axis. When $d_s=20$,
the model has the best overall performance across the datasets. This relates to the fact that the size of the character vocabulary used to generate the location hashing strings is 32. 
If $d_s$ is much smaller than 32, the character embedding may not  capture sufficient information from the hashing strings. Meanwhile, if $d_s$ is too large, the 
 information captured by each dimension may become too weak due to data sparsity. Another observation is that the model performs much better on Los Angeles. 
 This is because Los Angeles has the smallest number of POIs and average check-in sequence length per user. Its data  space to be learned is smaller 
 than those of the other two datasets.

\textbf{Impact of time context window size $k$.}
We vary the time context window size $k$ for our temporal pattern learning module from 1 to 5 
and report the results in Fig.~\ref{fig:ds_w}b. We see that SANST 
is also robust against the time context window size. We find the best time context window size to be strongly correlated to the time span and the average length of the check-in sequences in a dataset. 
In particular, Los Angeles covers a long time span (20 months) while its average check-in sequence length is the shortest. This means that 
its check-ins are less dense in the time dimension. Thus, it needs a larger time context window to achieve better performance, which explains for 
its increasing trend in nDCG@10 as $k$ increases towards 5. 

\textbf{Impact of model structure.}
The transformer network can be stacked with multiple layers, while the attention network can have multi-heads. 
We show the impact of the number of transformer network layers $\tau$ and the number of heads $h$ in Table~\ref{tab:model_structure}.
We see better results on Los Angeles with $\tau=2$, i.e., two transformer layers, which demonstrates that a deeper self-attention network may be helpful for learning the data patterns. 
However, deeper networks also have the tendency to overfit. The model performance drops when $\tau=2$ on the other two datasets. When $\tau=3$, the model is worse on all three datasets. 
We keep $\tau=2$ and further add more heads to the attention network. 
As the table shows, adding more heads does not help the model performance. This is opposed to observations 
in natural language processing tasks where different attention heads are more effective to capture various type of relations (e.g., positional and syntactic dependency relations)~\cite{voita-etal-2019-analyzing}. 
We conjecture that the relations between POIs are simpler (and without ambiguity) than those between words in natural language. Thus, a single-head attention is sufficient for our task.

\begin{table}[t]
\centering
\small
\caption{Impact of Model Structure (NDCG@10)} 
\label{tab:model_structure}
\begin{tabular}{lccc}
\toprule
Structure  & Gowalla & Los Angeles & Singapore \\
\midrule
$\tau$=1, $h$=1            & \textbf{0.1426}  & 0.2462      & \textbf{0.1539}    \\
$\tau$=2, $h$=1 (default) & 0.1374  & \textbf{0.2558}      & 0.1531    \\
$\tau$=3, $h$=1            & 0.1278 & 0.2416      & 0.1464    \\
\midrule
$\tau$=2, $h$=2 & 0.1358  & 0.2516      & 0.1464  \\
\bottomrule
\end{tabular}
\end{table}

\section{Conclusions}
We studied the next POI recommendation problem and proposed a self-attentive network 
named SANST for the problem. Our SANST model takes advantage of self-attentive networks 
for their capability in making sequential recommendations.  Our SANST model also incorporates 
the spatial and temporal patterns of user  check-ins.  As a result,  SANST retains the high 
efficiency of self-attentive networks  while enhancing their power in making recommendations 
for next POI visits. 
Experimental results on real-world datasets confirm the superiority of SANST -- it outperforms state-of-the-art sequential recommendation models and adapted baseline models that combine self-attentive networks with spatial features directly by up to 13.65\% in nDCG@10.

\bibliographystyle{named}
\bibliography{ijcai20}

\end{document}